\documentstyle[12pt]{article}
\begin{document}
\thispagestyle{empty}
\begin{flushright}
UA/NPPS-2-99
\end{flushright}
\vspace{3cm}

\begin{center}
{\large{\bf Extension of worldline computational algorithms for QCD\\
     to open fermionic contours}\\}

\vspace{1cm}

A. I. Karanikas and C. N. Ktorides\\
\smallskip
{\it University of Athens\\
Department of Physics\\
Nuclear and Particle Physics Section\\
Panepistimiopolis\\
GR-15771 Athens, Greece\\}
\end{center}

\vspace{2cm}

\begin{abstract}

The worldline casting of a gauge field system with spin-1/2 matter fields
has provided a, particle-based, first quantization formalism in the
framework of which the Bern-Kosower algorithms for efficient computations
in QCD acquire a simple interpretation. This paper extends the scope of
applicability of
the worldline scheme so as to include open fermionic paths. Specific
algorithms
are established which address themselves to the fermionic propagator and
which are directly applicable to any other process involving external
fermionic states. It is also demonstrated that in this framework the sole
agent of dynamics operating in the system is the Wilson line (loop)
operator, which makes a natural
entrance in the worldline action; everything else is associated with
geometrical properties of particle propagation, of which the most
important
component is Polyakov's spin factor. 

\end{abstract}

\newpage
\setcounter{page}{1}
{\large {\bf 1. Introduction}}

\vspace{.3cm}

 Consider a gauge theory with spin-1/2 matter fields, in the defining
representation of the symmetry group, described by a standard (renormalizable) 
Lagrangian density. A worldline transcription of such a system is
based, after a Wick rotation to Euclidean space-time, on a recasting
of the form $\int {\cal D}\overline{\psi}(x)\,{\cal D}\psi
(x)\,e^{S[\overline{\psi}(x),\psi (x), A_\mu (x)]}...\longrightarrow \int
{\cal D}x(\tau)\,{\cal D}p(\tau)\,e^{S[x(\tau),p(\tau),A_\mu(x(\tau))]}... $
according to which the matter fields register as point-like objects whose 
propagation (in phase-space) is furnished by a path integral. To the extent 
that the functional integration over the fermionic fields is Gaussian (in
the Grassmann sense), nothing is lost, in principle, from the content of the 
original system through the above transcription. Pioneering work that
dealt systematically with the subject appears in Refs [1-3].

More recent advancements in worldline methodology, especially in connection 
to gauge theories, have been made, independently, by Fried [4-6],
Strassler [7] and our group [8-12]. Respective methodological tools
adopted by the above 
authors are: (a) Schwinger's functional calculus formalism [13], (b) the 
string-inspired computational algorithms of Bern and Kosower [14] and
(c) Polyakov's description of the propagation of particle-like excitations
in (Euclidean) space-times [15]. Concerning specific objectives and/or 
applications aspired to by these attempts, the following rough picture
emerges. Fried and collaborators [4-6] as well as the present authors and
collaborators [10-12] have so far focused their efforts on eikonal-type 
descriptions in gauge field theories (with QCD as the ultimate target), 
along with associated issues such as
scattering amplitudes at asymptotically large collision energies,
factorization of soft physics, heavy quark effective theories, etc. In
Strassler's case, on the other hand, emphasis from the beginning was
placed on the development of new, more efficient rules for performing 
loop calculations in pQCD as per the realizations of Ref [14].

A key aspect of the Polyakov-inspired approach, that we have been advocating, is
its ability to accentuate geometrical features of particle-like propagation
in space-time. To illustrate this occurrence consider, first, the direct 
performance of the path integration over ${\cal D}p(\tau)$\footnote{In fact, 
the characterization ``worldline" strictly applies to the
resulting structure which, through this integration, formulates itself in
coordinate space.}. Adhering to the use of the first order form for
the Dirac operator we were able [9] to reproduce Polyakov's {\it spin
factor}
which characterizes the closed path propagation of a spin-1/2 particle-like
entity in Euclidean space-time. This quantity has a purely geometrical
content (it is, in some sense, associated with torsion) and accounts in a
self-consistent way, for the re-entry of a closed (Euclidean) space-time
contour by the spin-1/2 particle. By contrast, all other approaches to
worldline descriptions have adopted the second order form for the Dirac
operator which forces a, conventional type, spin-dependent term of the 
form $\sigma_{\mu\nu}F^{\mu\nu}$ in the worldline action. Our first
undertaking, in this paper, is to establish a direct connection between
Polyakov's geometric spin-factor and the ``dynamical" $\sigma \cdot F$
term.

A second consequence resulting from our approach
is its adaptation to configurations involving {\it open}
worldline contours. In physical terms this means that we are in a position to
extend our considerations to processes wherein spin-1/2 entities enter as
{\it external} particle states\footnote{It is in this context that common
grounds have been established with the eikonal-based work of Fried [4-6],
who
employs Schwinger's functional calculus as methodological tool and
explicitly accommodates spin through the $\sigma\cdot F$ term.}. By contrast,
Strassler's applications pertain to n-point functions with gauge field
modes as external states which attach themselves to a fermionic loop [7].
The 
major part
of our present effort is to extend worldline computational algorithms, 
hitherto provided for diagrams involving fermionic loops,
to ones wherein {\it open} fermionic paths make their entrance.

Our ultimate hope is that, through the present work, the elegant
separation between purely geometrical properties of matter particle paths, 
on the one hand, and dynamics, on the other, will emerge as one of the
most potent aspects of the worldline casting of gauge field theories.
Leaving the elucidation of this remark to an {\it a posteriori} discussion,
let us proceed to exhibit the organization of the present paper. In the
next section we display worldline expressions which will form the basis
of our subsequent considerations. We shall proceed, in Section 3, to make
the connection between the ``dynamical" $\sigma\cdot F$ term, appearing 
in the worldline action as a result of adopting the second order form for 
the Dirac operator and the ``geometrical"
Polyakov's spin factor. This is accomplished via a fortuitous application 
of an area derivative operator acting on functionals [16,17]. Section 4 
constitutes a central part of our paper. Specifically, we undertake the
task of carrying out the procedure that will lead to an explicit worldline
formula for the fermionic Green's function to second order in its
perturbative expansion. A major aspect of this attempt is to
present an expression which employs super-particle coordinates, i.e. a
path integral description which uses both bosonic and fermionic
variables. We proceed, in Section 5, to discuss
computational tools which are involved in practical applications of the
worldlike scheme. A first such undertaking is carried out in Section
6 where we establish, to second order, the equivalence between the Feynman
diagrammatic and worldline perturbative expansions of the fermionic
propagator in a non-abelian gauge theory. A direct one loop calculation is
performed in Section 7; it produces a final expression in parametrized
form which is ready to enter perturbative estimates of higher order
Green's functions with external fermion (as well as gauge field) legs. Our
concluding remarks are presented in Section 8, while some technical
matters pertaining to the work in Section 7 are displayed in an Appendix.

\vspace*{1cm}

{\large {\bf 2. Basic worldline expressions}}

\vspace{.3cm}

In this section we shall discuss worldline formulas placing emphasis on 
fermionic Green's functions in a background of, non-abelian in general, 
gauge fields. We shall work in Euclidean space-time in the context of which
the spin-factor can be defined. Special emphasis will be placed on the
second order Dirac formalism as our immediate aim (cf. next section) is
to establish its connection with the spin-factor.

Let us commence our considerations with a short presentation of previous
results stemming from the employment of the first order Dirac operator. 
A systematic procedure which takes one from the 
original field theoretical casting of the system to worldline formulas
can be found in [9]. Here we shall proceed formally. 
Consider, first, the effective action 
functional, with dependence on the background gauge field. Its
formal expression is

\begin{equation}
W[A]=ln\, Det[i\gamma\cdot D-(m-i\epsilon)],\,D_\mu=\partial_\mu +igA_\mu.
\end{equation}

Using Schwinger's proper time representation [18] and going to Euclidean
space-time we write
\begin{equation}
W[A]=-Tr\int_0^\infty\, \frac{dT}{T}e^{-T(-\gamma\cdot D+m)}.
\end{equation}

Computation of the non-Dirac $\gamma$ matrix as well as non-color part of
the trace in
coordinate space representation leads, after a suitable choice of 
gauge\footnote{We are here referring to a  gauge choice with respect to
parametrization invariance and mass identification. More specific comments
on this issue will
be made shortly.}, to the worldline result [9]
\begin{equation}
W[A]=-\int_0^\infty\, \frac{dT}{T}e^{-Tm}\int_{x(0)=x(T)}
{\cal D}x(\tau)\delta[\dot{x}(\tau)^2-1]tr\Phi (C)Tr_cPexp\left[ig
\int_0^T\dot{x}\cdot A(x(\tau))\right],
\end{equation}
where $tr(Tr_c)$ denotes trace over $\gamma$-matrix (color), $P$ stands
for path ordering and where the {\it spin-factor}
$\Phi(C)$ [15] is given by
\begin{equation}
\Phi(C)= Pexp\left\{\frac{1}{8}\int_0^T d\tau\,\omega_{\mu\nu}[x(\tau)]
[\gamma_\mu,\gamma_\nu]\right\}
\end{equation}
with 
\begin{equation}
\omega_{\mu\nu}(x(\tau))=\frac{1}{2}[\ddot{x}_\mu(\tau)\dot{x}_\nu(\tau)-
\dot{x}_\mu(\tau)\ddot{x}_\nu(\tau)],\quad \dot{x}^2=1,
\end{equation}
describing the orientation tensor of the local, perpendicular to the
path, plane.

As already mentioned, our main preoccupation in this paper will be with
fermionic Green's functions (propagators) whose definition involves open
paths. To begin, let us discuss worldline expressions for the
fermionic Green's function, in a (non-abelian) gauge field background. 
We first
consider the case where the first order form for the Dirac operator is
adopted. The corresponding Schwinger proper time formula for the
Green's operator is 
\begin{equation}
iG[A]=-\int_0^\infty dT\,e^{-T(-\gamma\cdot D+m)}.
\end{equation}
Its worldline representation, in phase-space, reads
\begin{eqnarray}
iG(x,y|A)&=&\int{\cal D}e(\tau)\int{\cal D}h(\tau)\int_{\stackrel 
{x(0)=x}{x(T)=y}}{\cal D}x(\tau)\int{\cal D}p(\tau)
exp[i\int_0^1d\tau h(\tau)\dot{e}(\tau)]\nonumber\\
& &\times exp[i\int_0^1 d\tau p(\tau)\cdot\dot{x}(\tau)]
P\,exp[-\int_0^1 d\tau e(\tau)\{m+i\gamma\cdot p(\tau)\}]\nonumber\\
& &\times P\,exp[ig\int_0^1A(x(\tau))\cdot\dot{x}(\tau)].
\end{eqnarray}
The integral over ${\cal D}h(\tau)$ on the right hand side supplies
our commitment to a choice of gauge with respect to the reparametrization 
invariance requirement. The latter, corresponds to the transformation
$\tau\rightarrow t=f(\tau),\, \dot{f}(\tau)>0,\,f(0)=0,\,f(1)=1$, under
which
we have $e(\tau)\rightarrow \dot{f}(\tau)e(t)$,
$\dot{x}(\tau)\rightarrow
\dot{f}(\tau)\dot{x}(t)$ and
$h(\tau)\rightarrow{1\over\dot{f}(\tau)}h(t)$.

The second order formalism for the Dirac operator, on the other hand,
bases itself on the
operator expression
\begin{equation}
iG[A]=(\gamma\cdot D+m)\int_0^\infty dT\,e^{-T[m^2-(\gamma\cdot D)^2]}
\end{equation}
whose phase-space worldline representation turns out to be

\begin{eqnarray}
iG(x,y|A)&=&\int{\cal D}e(\tau)\int{\cal D}h(\tau) \int_{\stackrel 
{x(0)=x}{x(1)=y}}{\cal D}x(\tau)\int{\cal D}p(\tau)
\left[m-{1\over 2e(1)}\gamma\cdot\dot{x}(1)\right]\nonumber\\
& &\times exp\left[i\int_0^1d\tau h(\tau)\dot{e}(\tau)\right]
exp[i\int_0^1 d\tau p(\tau)\cdot\dot{x}(\tau)]\nonumber\\
& &\times exp\left[-\int_0^1 d\tau e(\tau)\{p^2(\tau)+m^2\}\right]\nonumber\\
& &\times P\,exp\left[ig\int_0^1A(x(\tau))\cdot\dot{x}(\tau)+\frac{1}{2}g
\int_0^1d\tau e(\tau)\sigma\cdot F(x(\tau))\right].
\end{eqnarray}

As already stated our first preoccupation in this paper is to
establish that the $\sigma\cdot F$ term in the above formula translates
into the analogue of the spin-factor for {\it open} fermionic paths.  	 
The first step in  this direction is to 
carry out the integration over ${\cal D}p(\tau)$.

We obtain
\begin{eqnarray}
iG(x,y|A)&=&\int{\cal D}e(\tau)\int{\cal D}h(\tau)
exp\left[i\int_0^1d\tau h(\tau)\dot{e}(\tau)\right]\,C(e)\nonumber\\
& &\times\int_{\stackrel{x(0)=x}{x(1)=y}}{\cal D}x(\tau)
[m-{1\over 2e(1)}\gamma\cdot\dot{x}(1)]\nonumber\\
& &\times exp\left[-{1\over 4}\int_0^1 d\tau
\frac{\dot{x}^2(\tau)}{e(\tau)}-m^2\int_0^1 d\tau e(\tau)\right] 
\nonumber\\
& &\times P\,exp\left[ig\int_0^1A(x(\tau))\cdot\dot{x}(\tau)+\frac{1}{2}g
\int_0^1d\tau e(\tau)\sigma\cdot F(x(\tau))\right],
\end{eqnarray}
where
\begin{equation}
C(e)=\Pi_{\displaystyle {n=1}}^{\displaystyle{N}}\left(\frac{N}{4\pi e_n}
\right)^{D/2}
\end{equation}
can be looked upon as a normalization factor.

Our immediate objective is to recast the above expression into a form
where
the open-line analogue of the spin-factor replaces the $\sigma\cdot F$ term.
The relevant work will be carried out in the following section.

\vspace*{1cm}

{\large {\bf 3. Spin-factor in the second order formalism}}

\vspace{.3cm}

We commence our considerations which will lead to the recasting of the
$\sigma\cdot F$ term in (13) by utilizing the area derivative operator for
path
dependent functionals. We shall follow, in this respect, Polyakov's
proposal [16] for practical
reasons. A geometrical definition has been given by
Migdal [17].
Let us, then, introduce the operator (area derivative) by 
\begin{equation}
\frac{\delta}{\delta s_{\mu\nu}(t)}\equiv lim_
{\varepsilon
\rightarrow 0}\int^{+\varepsilon}_{-\varepsilon}d\sigma\, \sigma
\frac{\delta^2}{\delta x_\mu(t+\sigma/2)\delta x_\nu(t-\sigma/2)}.
\end{equation}

Consider the action of the above operator on the functional
\begin {equation}
I[\dot{x}(\tau)]\equiv exp\left[-{1\over 4}\int_0^1d\tau
\frac{\dot{x}^2(\tau)}
{e(\tau)}\right]
\end{equation}
which leads to
\begin{equation}
\frac{\delta}{\delta s_{\mu\nu}(t)}exp\left[-{1\over 4}\int_0^1d\tau 
\frac{\dot{x}^2(\tau)}  
{e(\tau)}\right]={1\over 2}\omega_{\mu\nu}(x(t))\left[-{1\over 4}
\int_0^1d\tau \frac{\dot{x}^2(\tau)}{e(\tau)}\right]
\end{equation}
with $\omega_{\mu\nu}$ now given by 
\footnote{Note that the expression that follows coincides with eq. (5)
when $e^2(\tau)=1$, which corresponds to a specific choice of gauge.}
\begin{equation}
\omega_{\mu\nu}(x(\tau))=\frac{1}{2e^2(\tau)}
[\ddot{x}_\mu(\tau)\dot{x}_\nu(\tau)-
\dot{x}_\mu(\tau)\ddot{x}_\nu(\tau)].
\end{equation}
On the other hand, we find

\begin{eqnarray}
& &P\,exp\left[ig\int_0^1A(x(\tau))\cdot\dot{x}(\tau)+\frac{1}{2}g
\int_0^1d\tau e(\tau)\sigma\cdot F(x(\tau))\right]=\nonumber\\
& &\quad = Pexp\left[-{i\over 2}\int_0^1d\tau e(\tau)\,\sigma \cdot
\frac{\delta}{\delta s(\tau)}\right]
P\,exp\left[ig\int_0^1d\tau\dot{x}(\tau) \cdot
A(x(\tau))\right].
\end{eqnarray}

Inserting into (10) we readily derive, after partial integration,
\begin{eqnarray}
iG(x,y|A)&=&\int{\cal D}e(\tau)\int{\cal D}h(\tau)  
exp\left[i\int_0^1d\tau h(\tau)\dot{e}(\tau)\right]\,C(e)\nonumber\\
& &\times\int_{\stackrel{x(0)=x}{x(1)=y}}{\cal D}x(\tau)
[m-{1\over 2e(1)}\gamma\cdot\dot{x}(1)]\nonumber\\
& &\times exp[-{1\over 4}\int_0^1 d\tau
\frac{\dot{x}^2(\tau)}{e(\tau)}-m^2\int_0^1 d\tau e(\tau)\nonumber\\
& &\times P\,exp\left[-{i\over 4}\int_0^1d\tau\,\sigma\cdot
\omega(x(\tau))\right] P\,exp\left[ig\int_0^1d\tau\dot{x}(\tau)\cdot 
A(x(\tau))\right].
\end{eqnarray}
Even though the term `spin-factor' {\it per se} pertains, as a geometrical
quantity, to closed paths, we shall employ, from hereon, this nomenclature
for the expression given by Eq. (4) even when it enters quantities defined
on open paths, such as the above.

The same considerations applied to the effective action functional whose
formal expression, in the second order formalism, is
\begin{equation}
-W[A]={1\over2}Tr\int_0^\infty\frac{dT}{T}e^{-T(m^2-(\gamma\cdot D)^2)},
\end{equation}
leads to the worldline representation
\begin{eqnarray}
-W[A]&=&\int{\cal D}e(\tau)\int{\cal D}h(\tau)
exp\left[i\int_0^1d\tau h(\tau)\dot{e}(\tau)\right]\,\bar{C}(e)\nonumber\\
& &\times\int_{x(0)=x(1)}{\cal D}x(\tau) exp[-{1\over 4}\int_0^1 d\tau
\frac{\dot{x}^2(\tau)}{e(\tau)}-m^2\int_0^1 d\tau e(\tau)\nonumber\\
& &\times tr P\,exp\left[-{i\over 4}\int_0^1d\tau\,\sigma\cdot
\omega(x(\tau))\right] Tr_cP\,exp\left[ig\int_0^1d\tau\dot{x}(\tau)
\cdot A(x(\tau))\right],
\end{eqnarray}
where $\bar{C}(e)\equiv \frac{1}{e(1)}C(e)$.

For a proper choice of gauge the above result matches the one that has been 
obtained [9] through the first order formalism, cf. Eq. (3). The notable 
accomplishment resulting from the manipulations that have just been
carried out is an apparent dissociation of a
geometrical feature (cf. spin-factor), characterizing the propagation of
particle-like modes, from the dynamics operating in
the system. The latter seem to be exclusively carried out by the
Wilson factor $P\,exp\left[ig\int_0^1d\tau\dot{x}(\tau)\cdot 
A(x(\tau))\right]$. As will actually turn out, 
computations involving the spin factor term do, in an implicit manner,
associate themselves with dynamical input.

Before proceeding further we shall commit ourselves to a choice of 
gauge in relation to reparametrization invariance. Designating as
dynamical
variable the proper time we set
\begin{equation}
t=\int_0^\tau d\tau'\,e(\tau'),\quad T=\int_0^1 d\tau\,e(\tau)
\end{equation}
and obtain
\begin{eqnarray}
iG(x,y|A)&=&\int_0^\infty dTe^{-Tm^2}\int_{\stackrel{x(0)=x}{x(T)=y}}
{\cal D}x(\tau)
[m-{1\over 2}\gamma\cdot\dot{x}(T)]
exp\left[-{1\over 4}\int_0^1 d\tau {\dot{x}^2(\tau)}\right]\nonumber\\
& &\times  P\,exp\left[-{i\over 4}\int_0^Td\tau\,\sigma\cdot
w(x(\tau))\right] P\,exp\left[ig\int_0^Td\tau\dot{x}(\tau)\cdot 
A(x(\tau))\right],
\end{eqnarray} 
where the measure ${\cal D}x(\tau)$ is now defined as
$lim_{N\rightarrow\infty}
\left(\frac{N}{4\pi T}\right)^{\frac{ND}{2}}\sum_{n=0}^Nd^Dx_n$, while 
$w_{\mu\nu}$ is, on account of our choice of gauge, given by
\begin{equation}
w_{\mu\nu}(x)={T\over 2}(\ddot{x}_\mu\dot{x}_\nu-\dot{x}_\mu\ddot{x}_\nu).
\end{equation}
Note that $|w|^2\sim |\ddot{x}|^2$, which means that it furnishes the curvature
when $|\dot{x}|=$const.

Turning our attention to the full fermionic Green's function we write
\begin{eqnarray}
iG(x,y)&=&\int_0^\infty dTe^{-Tm^2}\int_{\stackrel{x(0)=x}{x(T)=y}}{\cal
D}x(\tau)
\left[m-{1\over 2}\gamma\cdot\dot{x}(T)]
exp[-{1\over 4}\int_0^1 d\tau {\dot{x}^2(\tau)}\right]\nonumber\\
& &\times  P\,exp\left[-{i\over 4}\int_0^Td\tau\,\sigma\cdot
w(x(\tau))\right] <P\,exp\left[ig\int_0^Td\tau\dot{x}(\tau)\cdot
A(x(\tau))\right]>_A,
\end{eqnarray} 
where the expectation value in the gauge field sector includes gauge
fixing terms, ghost integration and Dirac determinant contribution. 

Generalizing, we infer that, in the worldline language, quantities of
theoretical
interest, such as n-point Green's functions and generating functionals,
draw all dynamical input from expectation values of Wilson lines or loops.
Such objects have been extensively studied in the past, see, for example,
[16, 19-23]. Any possible non-perturbative application of the present
formalism will, therefore, be
greatly facilitated by
relevant, accumulated knowledge.

Even though Eq. (23) will serve as the starting point for our subsequent
considerations, we find it useful to recast it in a form which adopts
particle-based coordinates. Along with practical consequences, this will
facilitate a direct comparison with
Strassler's (closed path) worldline expressions. We shall devote the
remainder of the present section to this task. 

As a first step, let us cast
(23) in the form $iG(x,y)=a(u^2,m^2) u\cdot\gamma+b(u^2,m^2), 
\,u_\mu\equiv {1\over T} (y-x)_\mu$ by writing
\begin{eqnarray}
iG(x,y)&=&{1\over D}\int_0^\infty
dTe^{-Tm^2}\int_{\stackrel{x(0)=x}{x(1)=y}}{\cal
D}x(\tau)\nonumber\\& &\times\left\{\left[{m\over D} \delta_{\alpha\beta}
-{1\over 2}\gamma_\alpha\dot{x}_\beta(T)\right]
tr\left[\gamma_\alpha\gamma_\beta  P\,exp\left[-{i\over
4}\int_0^Td\tau\,\sigma\cdot
w(x(\tau))\right]\right]\right\}\nonumber\\
& &\times <P\,exp\left[ig\int_0^Td\tau\dot{x}(\tau)\cdot
A(x(\tau))\right]>_A
\end{eqnarray}
which follows via the identifications $b={1\over D}\, tr(iG)$,
$au_\mu={1\over D}\,tr(\gamma_\mu iG)$ and once taking into account
that $trI=D$.

We now use the identity
\begin{eqnarray}
& &tr\left\{\gamma_\alpha\gamma_\beta  P\,exp\left[-{i\over
4}\int_0^Td\tau\,\sigma\cdot
w(x(\tau))\right]\right\} =\int_{\psi(0)
=\psi(T)=0}[d\psi]2\psi_\alpha(T)\psi_\beta(T)\nonumber\\& 
&\quad\quad\times exp\left[-{1\over2} \int_0^T
d\tau\psi\cdot\dot{\psi}+{1\over 2}\int_0^Td\tau\psi_\mu(\tau)
\psi_\nu(\tau)w_{\mu\nu}(x(\tau))\right]
\end{eqnarray}
where the $\psi_\mu$ are Grasmann variables whose correlator, with respect
to the action $\left[-{1\over2} \int_0^Td\tau\psi\cdot\dot{\psi}\right]$,
is determined by taking into consideration that
\begin{equation}
\{\psi_\mu,\psi_\nu\}=\delta_{\mu\nu}
\end{equation}
and
\begin{equation}
{1\over D}\int_{\psi(0)+\psi(T)=0}[d\psi]\,exp\left[-{1\over 2} \int_0^T
d\tau \psi\cdot\dot{\psi}\right]=1.
\end{equation}

We obtain
\begin{equation}
<\psi_\mu(\tau_1)\psi_\nu(\tau_2)>_\psi={1\over 2}\delta_{\mu\nu}
sign(\tau_1-\tau_2).
\end{equation} 

One is thereby led to the following particle-based, to be referred to as
`super-particle', representation for
the full fermionic propagator
\begin{eqnarray}
iG(x,y)&=&{1\over D}\int_0^\infty
dTe^{-Tm^2}\int_{\stackrel{x(0)=x}{x(1)=y}}{\cal
D}x(\tau)\left[{m\over D} \delta_{\alpha\beta}
-{1\over 2}\gamma_\alpha\dot{x}_\beta(T)\right]\int_{\psi(0)
+\psi(T)=0}[d\psi]\nonumber\\& &
\times 2\psi_\alpha(T)\psi_\beta(T)exp\left[
-{1\over 4}\int_0^Td\tau \dot{x}^2-{1\over2} \int_0^T
d\tau\psi\cdot\dot{\psi}+{1\over 2}\int_0^Td\tau\psi_\mu
\psi_\nu w_{\mu\nu}[x(\tau)]\right]\nonumber\\ & & \quad\times
 <P\,exp\left[ig\int_0^Td\tau\dot{x}(\tau)\cdot
A(x(\tau))\right]>_A.
\end{eqnarray}

This is an essentially new result which not only extends
Strassler's expressions to {\it open} fermionic lines but encodes spin
effects through a {\it geometrical quantity} entering a particle-based
action 
functional. At the same time it underlines the role of the Wilson
line operator as the sole agent of the dynamics operating in the system.
The spin
factor, in other words, has  joined the rest of the path-dependent (super)
coordinates to account for particle-based characteristics of spin-1/2 
matter field propagation. 
Generalizations of spin factor expressions pertaining to higher spins have
been discussed, from a different viewpoint, in Ref [24]. The bottom line
is that the spin factor accounts for geometrical features of paths
induced on them by the nature of the particle-like entity which 
traverses them.

Mention should be made, at this point, of the attempt in Ref [25] to
tackle the
problem of open line propagation of spin-1/2 particles in a gauge field
background. The relevant methodology has a different philosophy
(analytical rather
than geometrical) and restricts itself to situation wherein the gauge
field(s) do not acquire a
dynamical character, i.e. they remain external. By staying faithful to
Polyakov's geometric point of
view [15], on the other hand, we have both attained simplicity in form of
worldline expressions while achieving, at the same time, extensions beyond
the free particle case. 

The corresponding expression for the effective action functional is
\begin{eqnarray}
-W[A]&=&{1\over 2}\int_{\psi(0)
+\psi(T)=0}[d\psi]\int_0^\infty
dTe^{-Tm^2}\int_{\stackrel{x(0)=x}{x(T)=y}}{\cal
D}x(\tau)\nonumber\\& &\times 
exp\left[
-{1\over 4}\int_0^Td\tau \dot{x}^2-{1\over2} \int_0^T
d\tau\psi\cdot\dot{\psi}+{1\over 2}\int_0^Td\tau\psi_\mu
\psi_\nu w_{\mu\nu}(x(\tau))\right]\nonumber\\ & &\times Tr_c
P\,exp\left[ig\int_0^Td\tau\dot{x}(\tau)\cdot
A(x(\tau))\right].
\end{eqnarray}

The above equation corresponds to Strassler's result, modulo the presence
of the spin-factor
term in place of $\sigma\cdot F$. It serves as the
starting point for computational rules involving quark loops. The
relevant particle-based action can be read off the above
formulas and has as follows
\begin{equation}
S={1\over 4}\int_0^Td\tau \dot{x}^2+{1\over2} \int_0^T
d\tau\psi\cdot\dot{\psi}-{1\over 2}\int_0^Td\tau\psi_\mu
\psi_\nu w_{\mu\nu}(x(\tau))-ig\int_0^T d\tau\bar{z}\dot{x}\cdot A\,z,
\end{equation}
where the $z$-fields serve to enforce path ordering when background
fields are present.

Note that under the supersymmetric transformation $\delta x_\mu=
-i\theta\dot{\psi}_\mu$, $\delta\psi_\mu=\theta\dot{x}_\mu$ the first two
terms in (31) remain invariant whilst the last two do not. On the
other hand, we can easily retrieve the $\sigma\cdot F$ term, 
by partial integration, from the spin-factor. This
brings us right back to the Strassler form for the worldline action
which {\it is} supersymmetric invariant\footnote{Note that 
supersymmetry appears to be manifestly broken in (31) due to the
boundary conditions involved in the partial
integration which produced the spin factor.}. We shall refer to
(31) as super-worldline particle action.

\vspace*{1cm}

{\large {\bf 4. Analytic manipulations with the spin factor: Fermionic
propagator to 2nd order}}

\vspace{.3cm}

Besides its interpretational appeal as a geometrical agent that accounts
for the spin of a propagating particle, the spin factor presents
practical
advantages as it actually expedites analytic manipulations involved in
the computation of fermionic Green's functions and effective action
functionals. Samples of the latter case have been given in [9]. Here we
focus
our efforts on calculations pertaining to fermionic propagators.

\vspace{.6cm}

{\bf 4a. General Considerations}

\vspace{.2cm}

According to the resulting expressions in the previous section the spin
factor expression combines bosonic with fermionic particle-based
coordinates. Suppose we were to expand the relevant exponential. We would
then find ourselves having to compute expectation values in the bosonic
sector of the type $<(\dot{x}_\mu(\tau) \ddot{x}_\nu(\tau)
-\ddot{x}_\mu(\tau)\dot{x}_\nu(\tau))...>_x$. Clearly, the only way to get
a non-null result is for a four-vector, say $k_\mu$, to make its entrance
so
that, in combination with the four-vector $(x-y)_\mu$ which will emerge
from the path integration, would facilitate the creation of
antisymmetric combinations.
Equivalently, unless the first derivative is discontinuous at some 
$\tau=\tau_i$ the aforementioned expectation value terms, once recast the
form  
$<(\dot{x}_\mu(\tau) \dot{x}_\nu(\tau +\alpha)
-\dot{x}_\mu(\tau)\dot{x}_\nu(\tau-\alpha))...>_x\,(\alpha\rightarrow 0)$,
would lead to a vanishing result except at points where a
`force' acts, thereby `injecting' a four-momentum $k_\mu$ on the fermionic
line.

Consider the situation where $M$ such points $\tau_i$ are present on a
given fermionic line. Let us set
\begin{equation}
{1\over 2}[\dot{x}_\mu(\tau_i+\epsilon)-\dot{x}_\mu(\tau_i-\epsilon)]
=lim_{\epsilon\rightarrow 0}\,ik_{i\mu},\quad i=1,...,M.
\end{equation}

We surmise (here $C$ denotes an open curve)
\begin{eqnarray}
\Phi(C)&\rightarrow &lim_{\epsilon\rightarrow 0} \exp\left[{T\over 4}
\sum_{i=1}^{M}
\int_{\tau_i-\epsilon}^{\tau_i+\epsilon}d\tau \psi_\mu(\tau)
\psi_\nu(\tau)[\ddot{x}_\mu(\tau)\dot{x}_\nu(\tau)
-\ddot{x}_\nu(\tau)\dot{x}_\mu(\tau)\right]\nonumber\\ & &\quad
=\exp-{T\over 2}\sum_{i=1}^M
\psi_\mu(\tau_i)\psi_\nu(\tau_i)[ik_{i\mu}
\dot{x}_\nu(\tau_i)-ik_{i\nu}\dot{x}_\mu(\tau_i)].
\end{eqnarray}
Momentum conservation can be taken into account by setting
$\sum_{i=1}^Mk_{i\mu}=0$.

Clearly, $M$ measures the number of points where a gauge field is applied
on a given fermionic path\footnote{It is in this context that an earlier
comment, to
the effect that the computation of the spin-factor expression requires
dynamical input, was made.}. This means that $M$ coincides with the
perturbative order to which one expands the Wilson line in a given
worldline expression such as, e.g., (29).

Let us now return to the expansion of the exponential
that furnishes the spin factor, only this time we focus on its fermionic
component. From eq (26) we determine
\begin{equation}
<\psi_\mu(\tau)\psi_\nu(\tau)>_\psi={1\over2}\delta_{\mu\nu}.
\end{equation}
This means that any term in the expansion of order greater than $M$ will
inevitably lead to mixtures of symmetric and antisymmetric factors, hence
a vanishing contribution to the propagator. So, not only is $M$ associated
with perturbative order but also sets a bound on the contributing terms in
the expansion of the spin factor.

Armed with the above concrete as well as of practical value realizations,
we
are in position to turn our attention to the 2nd order expression of the
fermionic Green's function and gain specific insights with respect to spin
factor contributions to its worldline form. This is precisely the task
that we shall
carry out in the next subsection.

\vspace{.6cm}

{\bf 4b. Spin factor contribution to the second order fermionic Green's
function}
  
\vspace{.2cm}

Expanding the Wilson exponential in (29) we bring down, to second order in
$g^2$, the factor $\int_0^Td\tau_1\int_0^Td\tau_2\dot{x}_\mu(\tau_2)
\dot{x}_\nu(\tau_1)<A_\mu(x(\tau_2))A_\nu(\tau_1))>_A$. Working in the
Feynman gauge we write
\begin{eqnarray}
& &iG^{(2)}(x,y)=-g^2{1\over
D}c_F\int\frac{d^Dk}{(2\pi)^D}\,\frac{1}{k^2} 
\int_0^\infty
dTe^{-Tm^2}\int_0^Td\tau_1\int_0^Td\tau_2
\int_{\stackrel{x(0)=x}{x(T)=y}}{\cal D}x(\tau) \nonumber\\& &\quad
\times\dot{x}(\tau_2)\cdot\dot{x}(\tau_1)
\left[{m\over D} \delta_{\alpha\beta}
-{1\over 2}\gamma_\alpha\dot{x}_\beta(T)\right]
\int_{\psi(0)
+\psi(T)=0}[d\psi(\tau)]2\psi_\alpha(T)\psi_\beta(T)\nonumber\\ & &
\quad\times exp\left[
-{1\over 4}\int_0^Td\tau \dot{x}^2-{1\over2} \int_0^T
d\tau\psi\cdot\dot{\psi}+{1\over 2}\int_0^Td\tau\psi_\mu
\psi_\nu w_{\mu\nu}-ik\cdot(x(\tau_2)-x(\tau_1))\right].
\end{eqnarray}

Focusing our attention on the spin-factor expression, let us consider the
first order term, to be denoted by $\Phi^{(1)}$, in its
expansion. According to the general discussion of the previous subsection 
we determine
\begin{equation}
\Phi^{(1)}={T\over 2}
\sum_{i=1,2}\psi_\mu(\tau_i)\psi_\nu(\tau_i)[ik_{\mu i}
\dot{x}_\nu(\tau_i)-ik_{\nu i}\dot{x}_\mu(\tau_i)].
\end{equation}

Setting $k_2=-k_1=k$ we obtain
\begin{eqnarray}
\Phi^{(1)}&=&{T\over 2}\psi_\mu(\tau_2)\psi_\nu(\tau_2)[ik_{\mu}
\dot{x}_\nu(\tau_1)-ik_{\nu}\dot{x}_\mu(\tau_1)]\nonumber\\ & &
-{T\over 2}\psi_\mu(\tau_1)\psi_\nu(\tau_1)[ik_{\mu}
\dot{x}_\nu(\tau_2)-ik_{\nu}\dot{x}_\mu(\tau_2)].
\end{eqnarray}

Consider, next, the path integration over the $\psi$ fields. According to
(34) the contribution from the above terms is zero.
Had our computation referred to the effective action functional $W[A]$
this would be the end of the story, as far as path integration over the
fermionic coordinates is concerned. For the Green's function, on the other
hand, a four-spinor factor also appears for which we determine (we keep
only the antisymmetric contribution,
under the exchange $\mu\leftrightarrow\nu$)
\begin{equation}
2<\psi_\alpha(T)\psi_\beta (T)\psi_\mu(\tau_i)\psi_\nu(\tau_i)>_\psi
=-{1\over2}[\delta_{\alpha\mu}\delta_{\beta\nu} 
-\delta_{\alpha\nu}\delta_{\beta\mu}],\quad i=1,2.
\end{equation}
Reserving the same notation for the quantity that will
result
after the path integration with respect to the particle coordinates
we write 
\begin{equation}
\Phi^{(1)}\rightarrow{T\over 2}ik_\alpha[\dot{x}_\beta(\tau_2)-
\dot{x}_\beta(\tau_1)]\,-\,{T\over 2}ik_\beta[\dot{x}_\alpha(\tau_2)-
\dot{x}_\alpha(\tau_1)],
\end{equation}
where the arrow serves to signify the fact that a functional integration
with respect to the $x_\mu(\tau)$ fields remains to be carried out. 
Now,  along the way, the full computation of $\Phi^{(1)}$ will encounter
the quantity
$\dot{x}(\tau_2)\cdot\dot{x}(\tau_1)$. Taking into account that
\begin{equation}
<\dot{x}_\rho(\tau)\dot{x}_\sigma(\tau)>_x
={2\over T}\delta_{\rho\sigma},
\end{equation}
whose justification is, as we shall see later, intimately connected
with
reparametrization invariance, we are in position to replace (39) by
\begin{equation}
\Phi^{(1)}\rightarrow ik_\alpha[\dot{x}_\beta(\tau_1)-
\dot{x}_\beta(\tau_2)]\,-\,ik_\beta[\dot{x}_\alpha(\tau_1)-
\dot{x}_\alpha(\tau_2)].
\end{equation}

Next, we consider the contribution from the second order term
which, according to our general discussion, is given by
\begin{equation}
\Phi^{(2)}={T^2\over 8}\psi_\mu(\tau_2)\psi_\nu(\tau_2)
\psi_\kappa(\tau_1)\psi_\lambda(\tau_1)[k_\mu\dot{x}_\nu(\tau_2)-
k_\nu\dot{x}_\mu(\tau_2)][k_\kappa\dot{x}_\lambda(\tau_1)-
k_\lambda\dot{x}_\kappa(\tau_1)]+(1\leftrightarrow 2).
\end{equation}

Taking on the task of calculating expectation values in the $\psi$-sector
of the (super)particle system we determine, once keeping only the
symmetric term under the exchange $1\leftrightarrow 2$,
\begin{equation}
2<\psi_\alpha(T)\psi_\beta(T)\psi_\mu(\tau_2) \psi_\nu(\tau_2)
\psi_\kappa(\tau_1)\psi_\lambda(\tau_1)>_\psi={1\over 4}
\delta_{\alpha\beta}[\delta_{\mu\lambda}\delta_{\nu\kappa}-
\delta_{\nu\lambda}\delta_{\mu\kappa}][sign(\tau_1-\tau_2)]^2
\end{equation}

As before, we keep the notation $\Phi^{(2)}$ for the quantity which
results from path integration in the super-particle sector.
Substitution of (46) into (45) furnishes the intermediate result
\begin{eqnarray}
\Phi^{(2)}&\rightarrow &-{T^2\over 8}\delta_{\alpha\beta}
[k_\mu\dot{x}_\nu(\tau_2)-k_\nu\dot{x}_\mu(\tau_2)]
[k_\mu\dot{x}_\nu(\tau_1)-k_\nu\dot{x}_\mu(\tau_1)]G^2_F(\tau_1,\tau_2)
\nonumber\\& &\quad\quad = -{T^2\over 4}\delta_{\alpha\beta}
(k^2\delta_{\mu\nu}-k_\mu k_\nu)\dot{x}_\mu(\tau_1)\dot{x}_\nu(\tau_2)
G_F^2(\tau_1,\tau_2),
\end{eqnarray}
where we have, following Strassler, set 
$G_F(\tau_1,\tau_2)=sign(\tau_1-\tau_2)$.

Further utilization of the bosonic sector path integration allows us to 
cast $\Phi^{(2)}$ in the
more simplified form
\begin{equation}
\Phi^{(2)}\rightarrow -\delta_{\alpha\beta}
(k^2\delta_{\mu\nu}-k_\mu k_\nu)\delta_{\mu\nu}G_F^2(\tau_1,\tau_2) 
=-\delta_{\alpha\beta}(D-1)k^2 G_F^2(\tau_1,\tau_2).
\end{equation}

All higher order terms in the exponential expansion of the spin-factor
term give vanishing contribution to the Green's function. Here, we
explicitly witness the harmonization of the spin-factor with dynamical
aspects of the
calculation: Recognizing the order of the perturbative expansion the
spin-factor series
terminates when the number of vertices (points on the path where
a `force' is exerted) is saturated.

Putting together everything that was so far carried out, we arrive at the
following worldline expression for the fermionic propagator, to second
order in perturbation theory,
\begin{eqnarray}
& &iG^{(2)}(x,y)=-g^2
c_F\int\frac{d^Dk}{(2\pi)^D}\,\frac{1}{k^2}
\int_0^\infty
dTe^{-Tm^2}\int_0^Td\tau_1\int_0^Td\tau_2\theta(\tau_2-\tau_1)
\int_{\stackrel{x(0)=x}{x(T)=y}}{\cal D}x(\tau) \nonumber\\& &\quad
\times\left[{m\over D} \delta_{\alpha\beta}
-{1\over 2}\gamma_\alpha\dot{x}_\beta(\tau)\right]\{\delta_{\alpha\beta}
\dot{x}(\tau_2)\cdot\dot{x}(\tau_1)+ik_\beta[\dot{x}_\alpha(\tau_1)-
\dot{x}_\alpha(\tau_2)]-ik_\alpha[\dot{x}_\beta(\tau_1)-
\dot{x}_\beta(\tau_2)]\nonumber\\ 
& &\quad -\delta_{\alpha\beta}G^2_F(\tau_1,\tau_2)
(D-1)k^2\} exp\left[-{1\over 4}\int_0^Td\tau\,
\dot{x}^2-ik\cdot (x(\tau_1)-x(\tau_2))\right].
\end{eqnarray}

The corresponding expression for the effective action functional $W[A]$
follows in an analogous way and reads
\begin{eqnarray}
W[A]&=&{D\over 2}g^2\int\frac{d^Dk}{(2\pi)^D}\,Tr_cA_\mu(k)A_\nu(-k)
\int_0^\infty
dTe^{-Tm^2}\int_0^Td\tau_1\int_0^Td\tau_2\theta(\tau_2-\tau_1)
 \nonumber\\& &\times
\int_{x(0)=x(T)}{\cal D}x(\tau)[\dot{x}_\mu(\tau_1)\dot{x}_\nu(\tau_2)  
-G^2_F(\tau_1,\tau_2)(k^2\delta_{\mu\nu}-k_\mu k_\nu)
\nonumber\\ & &\times exp\left[-{1\over 4}\int_0^Td\tau\,
\dot{x}^2-ik\cdot (x(\tau_1)-x(\tau_2))\right].
\end{eqnarray}

The above results express the embodiment of spin factor contribution to
worldline
formulas. Clearly, the immediate question is whether (46) furnishes
the same perturbative expression for the propagator that
one extracts, conventionally, through Feynman diagrammatic, rather than
worldline, reasoning. That this is, indeed, the case will
be established in Section 6. More interesting, of course, is the
question regarding whether the formalism that has been developed in this
work can define
autonomous approaches to field theoretical computations or, at least,
offer practical advantages for higher order perturbative calculations for
processes
involving external fermionic fields. Setting this issue aside we shall,
in the next section,
turn our attention to the establishment of a set of rules
pertaining to expectation values of 
bosonic `field' products.

\vspace{1cm}

{\large {\bf 5. Computational tools in the super-particle representation}}

\vspace{.3cm}

The novel feature of the worldline formalism is the super-particle mode of
description which pervades its expressions for each given quantity of
physical interest. Our spin factor manipulations in the previous section
has mainly concentrated on aspects
surrounding path integration over fermionic coordinates. In the present
section we shall conduct a systematic discussion of the particle-based
(super)sector whose ultimate aim is to promote the practical side of the
worldline scheme.

Focusing on the bosonic sector of super-particle actions we realize that
its kinetic term has the conventional form $\int_0^Td\tau\dot{x}^2(\tau)$.
Corresponding propagators can be appropriately
defined.
For open lines the Green's function is denoted by
$\Delta(\tau,\tau')$. It
obeys the equation
\begin{equation}
\frac{\partial^2}{\partial\tau^2}\Delta(\tau,\tau'))=-\delta(\tau-\tau'),
\end{equation}   
with boundary conditions $\Delta(0,\tau')=\Delta(T,\tau')=0$.

For closed contours the situation demands closer inspection. As argued by
Strassler [7],  the relevant Green's function, to be denoted
by $G_B(\tau,\tau')$, obeys an equation of the form
\begin{equation}
{1\over{\cal C}}\frac{\partial^2}{\partial\tau^2}G_B(\tau,\tau'))
=\delta(\tau-\tau')-{1\over T},
\end{equation}
where $T$ is the length, according to a chosen parametrization, of the
(closed) contour. Following Strassler we adopt the convention which sets
${\cal C}=2$.

One obtains, respectively,
\begin{equation}
\Delta(\tau,\tau')=\frac{\tau(T-\tau')}{T}\theta(\tau'-\tau)+
\frac{\tau'(T-\tau)}{T}\theta(\tau-\tau'),\quad
\frac{\partial}{\partial\tau}
\Delta(\tau,\tau')= \theta(\tau'-\tau) -{\tau'\over T}
\end{equation}
and
\begin{equation}
G_B(\tau,\tau')=\frac{|\tau-\tau'|}{T}(T-|\tau-\tau'|),\quad
\frac{\partial}{\partial\tau} G_B(\tau,\tau')
=sign(\tau-\tau')-{2(\tau-\tau')\over T}.
\end{equation}

We also mention the relations
\begin{equation}
\frac{\partial}{\partial\tau}[\Delta(\tau,\tau_2)-\Delta(\tau,\tau_1)]={1\over
2}
\frac{\partial}{\partial\tau}[G_B(\tau,\tau_1)- G_B(\tau,\tau_2)],\quad
\frac{\partial^2}{\partial\tau\partial\tau'}\Delta(\tau,\tau')= {1\over 2}
\ddot{G}_B(\tau,\tau'),
\end{equation}
where the double dot on $G_B$ signifies second derivative with
respect to $\tau$. Finally, during our calculations we shall, following
Strassler, use the rules  $\dot{G}_B(\tau,\tau)=0$ and
$\dot{G}_B^2(\tau,\tau)=0$.
$\dot{G}_B(\tau,\tau')=-\dot{G}_B(\tau',\tau)$.

As a first application let us establish the validity of Eq.
(40). We readily determine
\begin{eqnarray}
<\dot{x}_\rho(\tau)\dot{x}_\sigma(\tau)>_x&=&lim_{\epsilon\rightarrow 0}\,
{1\over T}\delta_{\rho\sigma}\int^{\tau-\epsilon}_{\tau+\epsilon}d\tau'
\ddot{G}_B(\tau\tau')\nonumber\\ &=&lim_{\epsilon\rightarrow 0}\,{2\over
T}\delta_{\rho\sigma}\int^{\tau-\epsilon}_{\tau+\epsilon}d\tau'\left
[\delta(\tau-\tau')-{1\over T}\right ]={2\over T}\delta_{\rho\sigma},
\end{eqnarray}
One, thereby, explicitly verifies that integration over the bosonic
coordinates $x(\tau)$ produces a result induced by parametrization
invariance as it yields a constant measure for the velocity:
$<|\dot{x}|^2>=\frac{2D}{T}$.

Next, we undertake the computation of expectation values of the
first few products of derivatives of bosonic `fields'
$\dot{x}_\mu(\tau)$, with action functional the one that arose in
connection with our second order fermionic propagator expression,
cf. eq. (46). The particle-based (bosonic) action
reads
\begin{equation}
S={1\over 4}\int_0^Td\tau\dot{x}^2\,+\,ik\cdot[x(\tau_2)-x(\tau_1)].
\end{equation}

The fact that terms of higher than 2nd power are absent in $S$ encourages
the employment of techniques which base themselves on solutions of the
classical
equations of motion. The latter read 
\begin{equation}
\ddot{x}_\mu^{cl}(\tau)\,=\,2ik_\mu[\delta(\tau-\tau_2)-
\delta(\tau-\tau_1)].
\end{equation}
with solution
\begin{equation}
x_\mu^{cl}(\tau)\,=\,\eta_\mu(\tau)-2ik_\mu[\Delta(\tau,\tau_2)-
\Delta(\tau,\tau_1)],
\end{equation}
where $\eta_\mu(\tau)=\frac{(y-x)_\mu}{T}\tau+x_\mu=u_\mu\tau+x_\mu$.

Under the substitution $x(\tau)\rightarrow x(\tau)+x^{cl}(\tau)$ the
action becomes
\begin{equation}
S\,=\,S_{cl}+{1\over 4}\int_0^Td\tau\dot{x}^2(\tau),
\end{equation}
with
\begin{equation}
S_{cl}={1\over 4}Tu^2+ik\cdot u(\tau_2-\tau_1)+k^2G_B(\tau_1,\tau_2),
\end{equation}

Consider, first, the case of $<\dot{x}_\mu(\tau)>$. We readily determine
\begin{equation}
<\dot{x}_\mu(\tau)>={\cal N}^{-1}\int_{\stackrel{x(0)=x}{x(T)=y}}{\cal D}
x(\tau) \dot{x}_\mu(\tau)e^{-S}=\dot{x}_\mu^{cl}(\tau),
\end{equation}
where we have defined
\begin{equation}
{\cal N}=\int_{\stackrel{x(0)=x}{x(T)=y}}{\cal D} x(\tau) e^{-S}.
\end{equation}

Note that
\begin{eqnarray}
\dot{x}_\mu^{cl}(\tau)&=&u_\mu-2ik_\mu\frac{\partial}{\partial\tau}
[\Delta(\tau,\tau_2)
- \Delta(\tau,\tau_1)]\nonumber\\& &=u_\mu-ik_\mu[\dot{G}_B(\tau,\tau_1)
-\dot{G}_B(\tau,\tau_2)].
\end{eqnarray}
which, given the equal time rules mentioned after eq (52) along with the
fact that $\dot{G}_B(\tau,\tau')=-\dot{G}_B(\tau',\tau)$, yields
\begin{equation} 
\dot{x}_\mu^{cl}(\tau_i)=u_\mu+ik_\mu\dot{G}_B(\tau_1,\tau_2),\quad i=1,2.
\end{equation}
Observe, in passing, the consistency of the above result with substitution
(32) associated with non-vanishing contributions of the spin factor in the
bosonic sector. Indeed, we determine
\begin{equation} 
<\dot{x}_\mu(\tau_1+\epsilon)- \dot{x}_\mu(\tau_1-\epsilon)>_x =
-ik_\mu[\dot{G}_B(\tau_1+\epsilon_1,\tau_1)-
\dot{G}_B(\tau_1-\epsilon_1,\tau_1)] =-2ik_\mu=2ik_{1\mu}.
\end{equation}

For the correlator we find
\begin{eqnarray}
<\dot{x}_\mu(\tau_1) \dot{x}_\mu(\tau_2)>_x&=&\dot{x}_\mu^{cl}(\tau_1)
\dot{x}_\mu^{cl}(\tau_2)+{\cal N}_o\int_{x(0)=x(T)=0}{\cal D}x(\tau) 
\dot{x}_\mu(\tau_1)
\dot{x}_\nu(\tau_2)e^{-{1\over 4}\int_0^Td\tau\dot{x}^2}\nonumber\\&
&\quad
=\dot{x}_\mu^{cl}(\tau_1)
\dot{x}_\mu^{cl}(\tau_2)+\delta_{\mu\nu}\ddot{G}_B(\tau_1,\tau_2),
\end{eqnarray}
where ${\cal N}_o=\frac{1}{(4\pi T)^{D/2}}$.
 
The last relation brings to surface the fact that
$\ddot{G}_B(\tau_i,\tau_j)$ will make regular appearances into our
expressions. An immediate problem can be identified with its delta
function component whose presence becomes the source of potential
ultraviolet
problems. The way to confront this issue is by performing an
integration by parts which replaces $\ddot{G}_B(\tau_i,\tau_j)$ by
$\dot{G}_B^2(\tau_i,\tau_j)$. Even at that one could still anticipate
problems associated with the end point behavior of integrations with
respect to the $d\tau_i$'s, where $\tau_i$ denotes a point of gauge field
contact with the super-particle contour. As with the case of closed
contours [7], what saves the situation is the fact that partial
integration
invariably produces the combination
$\dot{G}_B^2(\tau_i,\tau_j)-G_F^2(\tau_i,\tau_j)$ which eliminates any
pathological behavior\footnote{As an explicit illustration of this
occurrence the reader is referred to the next section 
where a computation pertaining to $G^{(2)}(x,y)$ will be carried out.}. The
super-particle description, which, in our treatment, has been introduced
through the spin factor, plays a crucial role in producing singularity
free expressions. 

In conclusion, our work in this section underlines the parallelism between
open and closed line computational subtlelties. Nothing that arises in the
former fails to do so in the latter case, or vice versa. The only
difference is that additional terms in
the super-particle action are
induced by the new boundary conditions but which have no fundamental 
impact, as far as
divergent behavior is concerned. Even though we shall not, in the present
paper, study physical quantities in which external gauge field lines
attach themselves onto matter particle contours we fully expect to
encounter analogous computational features, e.g. pinch singularities, in
these situations as well. We plan to report on such
calculations in the near future. 

\vspace*{1cm}

{\large {\bf 6. First application: Recovery of Feynman diagrammatic
expansion to 2nd order}}

\vspace{.3cm}

We have already pointed out that nothing is lost to the original content
of the field theoretical system on the account of manipulations that
have taken place for the purpose of casting it into its worldline form. As
a first illustration -which, at the same time, will enable us to see the
corresponding particle-based computational tools at work- we intend, in
this section, to prove the equivalence between the worldline and
Feynman perturbative expansions for the fermionic propagator, to second
order. Given the notable calculational advantages that have already been
established for non-abelian gauge field theories in connection with
closed particle contours [14, 7, 26], one is encouraged to think that our
present
extension will lead to efficient ways for perturbative estimates, to a
given order, for processes involving open matter particle contours as
well.

Let us commence our considerations by dispensing with the 0th order term
of the fermionic propagator. As per our discussion in the previous
section, we expect no contribution from the spin factor in the super
particle action. Indeed, the only available four-vector in the free
propagation case is $u_\mu=\frac{1}{T}(x-y)_\mu$ and there is no way to
generate antisymmetric combinations from it. The relevant worldline
expression reads
\begin{equation}
iG^{(0)}(x,y)=\int_0^\infty
dTe^{-Tm^2}\int_{\stackrel{x(0)=x}{x(T)=y}}{\cal D}
x(\tau)[m-{1\over 2}\gamma\cdot\dot{x}(T)]e^{-{1\over
4}\int_o^Td\tau\dot{x}^2
(\tau)}.
\end{equation}

The solution of the classical equations of motion is, simply,
$x^{cl}(\tau)=\eta_\mu$, with $\eta_\mu$ as in eq. (56). Under the
substitution $x(\tau)\rightarrow x^{cl}(\tau)
+x(\tau)$ path integration is performed with the same action but with
boundary conditions $x(0)=x(T)=0$. Given that the classical value of the
action is ${1\over 4}T
u^2$ and in view of (59) we finally obtain
\begin{eqnarray}
iG^{(0)}(x,y)&=&\int_0^\infty dT\frac{e^{-Tm^2}}{(4\pi T)^{D/2}}[m-{1\over
2}
\gamma\cdot u]e^{-{1\over 4}Tu^2}\nonumber\\& &=\int_0^\infty
dT\,e^{-Tm^2}\int\frac
{d^Dq}{(2\pi)^D}e^{-iq(x-y)-Tq^2}(m-i\gamma\cdot q).
\end{eqnarray}
Integration over $dT$ finally produces the free fermionic propagator.

Focusing our attention to the mass containing term $I_m$ in
$iG^{(2)}(x,y)$ we first recall
that spin factor contributions are already contained in (46). Moreover, we
have carried out, in the previous section, the computation of
relevant expectation values with respect to
the bosonic action given by eq. (54). Putting everything, that has already
been worked out, together we
obtain
\begin{eqnarray}
I_m&=&-mg^2c_F\int\frac{d^Dk}{(2\pi)^D}{1\over k^2}\int_0^\infty\frac{dT}
{(4\pi
T)^{D/2}}e^{-Tm^2}\int_0^Td\tau_1\int_0^Td\tau_2\theta(\tau_2-\tau_1)
\nonumber\\& &\quad\times\left\{D\ddot{G}_B(\tau_1,\tau_2)+[u+ik\dot{G}_B
(\tau_1,\tau_2)]^2-(D-1)k^2G_F^2(\tau_1,\tau_2)\right\}\nonumber\\& &\quad
\quad
exp\left[-ik\cdot u(\tau_2-\tau_1)-k^2G_B(\tau_1,\tau_2)-{T\over
4}u^2\right].
\end{eqnarray}

Our immediate task is to perform an integration by parts in order to get
rid
of the undesirable quantity $\ddot{G}_B$. We readily determine
\begin{eqnarray}
& & \int_0^Td\tau_1\int_0^Td\tau_2\theta(\tau_2-\tau_1)
\ddot{G}_B(\tau_1,\tau_2)
exp\left[-ik\cdot
u(\tau_2-\tau_1)-k^2G_B(\tau_1,\tau_2)\right]\nonumber\\& &=
\int_0^Td\tau_2\dot{G}_B(\tau_2,0)
exp\left[-ik\cdot u\tau_2-k^2G_B(\tau_2,0)\right]+
\int_0^Td\tau_1\int_0^Td\tau_2
\theta(\tau_2-\tau_1)\nonumber\\& &\quad\times [k^2\dot{G}_B^2(\tau_1, 
\tau_2)
-ik\cdot u\dot{G}_B(\tau_1,\tau_2)] exp\left[-ik\cdot
u(\tau_2-\tau_1)-k^2G_B(\tau_1,\tau_2)\right].
\end{eqnarray}
For the reader who is acquainted with closed-contour worldline
manipulations the analogy with the above practice is familiar; simply
the exponential has more terms for open lines. The important point is that
in both situations the integration by parts goes through on the basis of
the adopted regularization, namely $\dot{G}_B(\tau,\tau)=0$.

Making the variable change $\tau_2-\tau_1=\alpha T$ and
$\tau_2+\tau_1=2\beta T$
we readily determine
\begin{eqnarray}
I_m&=&-mg^2c_F\int\frac{d^Dk}{(2\pi)^D}{1\over k^2}\int_0^\infty\frac{dT}
{(4\pi T)^{D/2}}T^2e^{-Tm^2}\int_0^1d\alpha\nonumber\\& &\times\left\{
(1-\alpha)\left[-4(D-1)k^2\alpha(1-\alpha)-(D-2)ik\cdot
u(1-2\alpha)+u^2+\frac{2D}
{T}\right]-{D\over T}\right\}\nonumber\\& &\times\quad\quad 
exp\left[-ik\cdot uT\alpha-
k^2T\alpha(1-\alpha)-{T\over 4}u^2\right].
\end{eqnarray}
Some comments are in order with respect to what has taken place in 
producing the
above result. The most obvious one is that $\alpha$ assumes the role of a
Feynman parameter as per the standard diagrammatic manipulation. Given
its introduction the following practical rules of substitution, with
regard to the original worldline expression, emerge
\begin{equation}
\dot{G}_B(\tau_1,\tau_2)=1-2\alpha,\quad
\dot{G}_B^2(\tau_1,\tau_2)-\dot{G}_F^2(\tau_1,\tau_2)=-4\alpha(1-\alpha).
\end{equation}
The last relation is especially notable as it explicitly underlines the
absence of divergencies in the integral over the Feynman parameter.

Turning our attention to the $\gamma$-matrix part, $I_\gamma$, of the
second order propagator we need the following results pertaining to
expectation values 
\begin{equation}
<\gamma\cdot\dot{x}(T)>=\frac{1}{(4\pi T)^{D/2}}
\gamma\cdot\dot{x}^{cl}(T),
\end{equation}
\begin{equation}
<k\cdot\dot{x}(T)\gamma\cdot\dot{x}(\tau_i)>=\frac{1}{(4\pi T)^{D/2}} 
[\gamma\cdot k\ddot{G}_B(\tau_i,T)+k\cdot\dot{x}^{cl}(T)
\gamma\cdot\dot{x}^{cl}(\tau_i)],\quad i=1,2,
\end{equation}
\begin{equation}
<\dot{x}(T)\cdot\dot{x}(\tau_i)>=\frac{1}{(4\pi T)^{D/2}}
[D\ddot{G}_B(\tau_i,T)+\dot{x}^{cl}(T)\cdot\dot{x}^{cl}(\tau_i)],\quad
i=1,2
\end{equation}
and
\begin{eqnarray}
<\gamma\cdot\dot{x}(T)\dot{x}^{cl}(\tau_1)\cdot\dot{x}^{cl}(\tau_2>&=&
\frac{1}{(4\pi
T)^{D/2}}[D\gamma\cdot\dot{x}^{cl}(T)\ddot{G}_B(\tau_1,\tau_2)+
\gamma\cdot\dot{x}^{cl}(\tau_1)\ddot{G}_B(\tau_2,T)\nonumber\\& &+
\gamma\cdot\dot{x}^{cl}(\tau_2)\ddot{G}_B(\tau_1,T)+
\gamma\cdot\dot{x}^{cl}(T)\dot{x}^{cl}(\tau_1))\cdot\dot{x}^{cl}(\tau_2)].
\end{eqnarray}

As before, we must integrate by parts and make the variable change which
introduces the Feynman parameter $\alpha$. Doing so we are led to the
following
result, in momentum space,
\begin{eqnarray}
\tilde{I}_\gamma(q)&=&ig^2c_F\int\frac{d^Dk}{(2\pi)^D}{1\over
k^2}\int_0^\infty dTT^2\int_0^1d\alpha(1-\alpha)exp[-T(m^2+q^2-2q\cdot k
+\alpha k^2)]\nonumber\\& &\times\{\gamma\cdot q[(D+2)m^2+(D-2)q^2
-2(D-2)q\cdot k]+\gamma\cdot k(D-2)(m^2+q^2)\}.
\end{eqnarray}

Combining (69) and (75) we finally obtain
\begin{eqnarray}
\tilde{G}^{(2)}&=&-g^2c_F\frac{1}{(q^2+m^2)^2}\int\frac{d^Dk}{(2\pi)^D}
{1\over k^2}\frac{1}{(q-k)^2+m^2}\{m[Dm^2+(D-4)q^2-2(D-2)q\cdot k]
\nonumber\\& &\quad -i\gamma\cdot q[(D+2)m^2+(D-2)q^2-2(D-2)q\cdot k]
-i\gamma\cdot
k(D-2)(m^2+q^2)\}\nonumber\\& &=\frac{1}{m-i\gamma\cdot q}\left[
-g^2c_F\int\frac{d^Dk}{(2\pi)^D}
{1\over k^2}\gamma_\mu\frac{m-i(\gamma\cdot q-\gamma\cdot k)}{(q-k)^2+m^2}
\gamma_\mu\right]\frac{1}{m-i\gamma\cdot q}
\end{eqnarray}
which explicitly shows that the worldline perturbative, to second order,
term for the fermionic propagator coincides with the one obtained via the
Feynman diagrammatic expansion.

\vspace*{1cm}

{\large {\bf 7. One loop calculation in the fermionic propagator}}

\vspace{.3cm}

In this section we shall apply the worldline approach to a loop
calculation pertaining to an open line which extends a recent closed
line result by Schubert [27], see also Ref [28]. Specifically, by going
back to
(35)
we shall perform the loop integration, equivalently, the integration over
$d^Dk$, thereby extracting an expression for the second order contribution
to the fermionic propagator in parametric form. So as not to bore the
reader with endless manipulations we shall restrict our attention to the
mass term entering (35) which reads
\begin{eqnarray}
& &I_m=-g^2c_F\int\frac{d^Dk}{(2\pi)^D}\,\frac{1}{k^2}
\int_0^\infty
dTe^{-Tm^2}\int_0^Td\tau_1\int_0^Td\tau_2\theta(\tau_2-\tau_1)
\int_{\stackrel{x(0)=x}{x(T)=y}}{\cal D}x(\tau) \nonumber\\& &\quad\times 
\{\dot{x}(\tau_2)\cdot\dot{x}(\tau_1)-G^2_F(\tau_1,\tau_2)
(D-1)k^2\} exp\left[-{1\over 4}\int_0^Td\tau\,
\dot{x}^2-ik\cdot (x(\tau_1)-x(\tau_2))\right].
\end{eqnarray}

Introducing a small mass $\lambda$ for infrared protection, as well as a
mass $\mu$ for the purposes of dimensional regularization we perform the
$k$-integrations making use of the formulae
\begin{equation}
\mu^{4-D}\int\frac{d^Dk}{(2\pi)^D}\frac{(k^2)}{k^2+\lambda^2}e^{-ik\cdot
x}=
\frac{\mu^{4-D}}{(4\pi)^{D/2}}\int_0^\infty
dL\,L^{-\frac{D}{2}}e^{-\lambda^2L}
\left(\frac{D}{2L}-\frac{x^2}{4L^2}\right)e^{-\frac{x^2}{4L}},
\end{equation}
where the parentheses around $k^2$ serve to point out that for numerator
equal to 1 the terms in parentheses inside the integral on the right hand
side are also substituted by 1.

Isolating the path integral component of $I_m$, i.e.
\begin{eqnarray}
{\cal I}&=&\int_{\stackrel{x(0)=x}{x(T)=y}}{\cal D}x(\tau)
\left[\dot{x}(\tau_2)\cdot\dot{x}(\tau_1)
+\frac{D-1}{4L^2}G^2_F(\tau_1,\tau_2)(x(\tau_2)-x(\tau_1))^2-
\frac{(D-1)D}{2L}G^2_F(\tau_1,\tau_2)\right]\nonumber\\& &\quad\times
 exp\left[-{1\over 4}\int_0^Td\tau\,
\dot{x}^2-\frac{1}{4L} (x(\tau_1)-x(\tau_2))^2\right],
\end{eqnarray}
we find ourselves having to calculate expectation values of bosonic
super-particle field coordinates with respect to the action
\begin{equation}
S={1\over 4}\int_0^Td\tau\dot{x}^2+{1\over L}(x(\tau_2)-x(\tau_1))^2.
\end{equation}

Leaving, for what follows, a number of technical details to the Appendix
we present the equations of motion in the form
\begin{equation}
\ddot{x}_\mu^{cl}(\sigma)={1\over L}\int_0^Td\sigma' B(\sigma' ,\sigma)
x_\mu^{cl}(\sigma'),
\end{equation}
where
\begin{equation}
B(\sigma_1,\sigma_2)=[\delta(\sigma_1-\tau_1) -\delta(\sigma_1-\tau_2)]
[\delta(\sigma_2-\tau_1) -\delta(\sigma_2-\tau_2].
\end{equation}

Next, we introduce the Green's function $\Delta^{(1)}(\sigma_1,\sigma_2)$
by
\begin{equation}
\int_0^Td\sigma[\delta(\sigma_1-\sigma)\frac{\partial^2}{\partial\sigma^2}-
{1\over
L}B(\sigma,\sigma_1)]\Delta^{(1)}(\sigma,\sigma_2)=-\delta(\sigma_1
-\sigma_2),
\end{equation}
obeying the boundary conditions
$\Delta^{(1)}(0,\sigma)=\Delta^{(1)}(T,\sigma)=0$.

The solutions to the classical equations of motion are readily determined
as
\begin{equation}
x_\mu^{cl}(\sigma)=\eta_\mu(\sigma)-{1\over L}(\tau_2-\tau_1)u_\mu
[\Delta^{(1)}(\sigma,\tau_2)-\Delta^{(1)}(\sigma,\tau_1)],
\end{equation}
with $\eta_\mu(\sigma)$ as given in (56).

The corresponding expression for the classical action is
\begin{equation}
S^{cl}={T\over 4}u\cdot\dot{x}^{cl}(T)=\frac{Tu^2}{4}+{1\over 4}u^2
\frac{(\tau_2-\tau_1)^2}{L+G_B(\tau_1,\tau_2)}.
\end{equation}

Making the substitution $x\rightarrow x+x^{cl}$ we have that
\begin{eqnarray}
& &<\dot{x}(\tau_1)\dot{x}(\tau_2)>\rightarrow
<\dot{x}(\tau_1)\dot{x}(\tau_2)>_o
+\dot{x}^{cl}(\tau_1)\dot{x}^{cl}(\tau_2)\nonumber\\
& &<(x(\tau_2)-x(\tau_1))^2>\rightarrow <(x(\tau_2)-x(\tau_1))^2>_o
+(x^{cl}(\tau_2)-x^{cl}(\tau_1))^2,
\end{eqnarray}
where
\begin{equation}
<A>_o\equiv M_o^{-1}\int_{x(0)=(x(T)=0}{\cal D}x(\tau)Ae^{-S[x]}
\end{equation}
with $S[x]$ as in (80) and $M_o^-1$ the usual renormalization factor.

From (84) we readily determine
\begin{equation}
\dot{x}^{cl}(\tau_1) =\dot{x}^{cl}(\tau_2)=u_\mu +{1\over
2}u_\mu(\tau_1-\tau_2)
\frac{\dot{G}_B(\tau_1,\tau_2)}{L+G_B(\tau_1,\tau_2)}
\end{equation}
and
\begin{equation}
[x^{cl}(\tau_2)-x^{cl}(\tau_1)]^2=(\tau_2-\tau_1)^2u^2\left(\frac{L}{L+G_B
(\tau_1,\tau_2)}\right)^2.
\end{equation}

In the Appendix we establish that
\begin{equation}
{\cal 
N}^{-1}<\dot{x}(\tau_1)\dot{x}(\tau_2)>_o=D\ddot{G}_B(\tau_1,\tau_2)-{D\over
2}\frac{\dot{G}^2_B(\tau_1, \tau_2)}{L+G_B(\tau_1,\tau_2)}
\end{equation}
and
\begin{equation}
{\cal N}^{-1}<(x(\tau_2)-x(\tau_1))^2>_o=
2D\frac{LG_B(\tau_1,\tau_2)}{L+G_B(\tau_1,\tau_2)},
\end{equation}
where ${\cal N}$ is defined in the Appendix.

Putting everything together we obtain (we suppress the $\tau_1,\tau_2$ of
the $G$'s for notational economy)
\begin{eqnarray}
I_m&=&-mg^2\frac{c_F}{(4\pi)^D}\mu^{4-D}\int_0^\infty dLe^{-\lambda^2L}
\int_0^\infty dT T^{-D/2}e^{-Tm^2}\int_0^T d\tau_1\int_0^T d\tau_2
\frac{\theta(\tau_2-\tau_1)}{(L+G_B)^{D/2}}\nonumber\\& &\times
\left\{D\ddot{G}_B-{D\over 2}\frac{\dot{G}_B^2}{L+G_B}+u^2\left[1+{1\over
2}
(\tau_2-\tau_1)\frac{\dot{G}_B}{L+G_B}\right]^2+\frac{D-1}{4L^2}G_F^2\left[2D
\frac{LG_B}{L+G_B}+\right.\right.\nonumber\\& &\left.\left.
+(\tau_2-\tau_1)^2u^2\frac
{L^2}{(L+G_B)^2}-\frac{(D-1)D}{2L}G_F^2\right]\right\} 
exp\left[-\frac{Tu^2}{4}-
\frac{u^2}{4}\frac{(\tau_2-\tau_1)^2}{L+G_B}\right].
\end{eqnarray}

After an integration by parts, which rids us of the presence of
$\ddot{G}_B(\tau_1,
\tau_2)$ in favor of the combination $\dot{G}_B^2-G_F^2$, we get, once 
performing
the substitution $\tau_2-\tau_1=\alpha T,\,\tau_1+\tau_2=2\beta
T,\,L\rightarrow LT$, 
\begin{eqnarray}
I_m&=&-mg^2\frac{c_F}{(4\pi)^D}\mu^{4-D}\int_0^\infty dL
\int_0^\infty dT T^{3-D}e^{-T(m^2+\lambda L)}\int_0^1 d\alpha\frac{1
-\alpha}{[L+\alpha(1-
\alpha)]^{D/2}}\nonumber\\&
&\times\left[u^2(D-1)\frac{\alpha^3(1-\alpha)}{[L+
\alpha(1-\alpha)]^2}-\frac{2D(D-1)}{T}\frac{\alpha(1-\alpha)} 
{L+\alpha(1-\alpha)}
+\frac{(D-2)u^2}{2}\frac{\alpha(1-2\alpha)}{L+\alpha(1-\alpha)}\right.
\nonumber\\& &\left. +u^2+\frac{D}{T}
\frac{1-2\alpha}{1-\alpha}\right]exp\left[-\frac{Tu^2}{4}- 
\frac{Tu^2}{4}\frac{\alpha^2}{L+\alpha(1-\alpha)}\right].
\end{eqnarray}

The corresponding expression for the Fourier transform reads
\begin{eqnarray}
\tilde{I}_m&=&-mg^2\frac{c_F}{(4\pi)^D}\mu^{4-D}\int_0^\infty dL
\int_0^\infty dT T^{3-D/2}\int_0^1
d\alpha\frac{1-\alpha}{(L+\alpha)^{D/2}}
\nonumber\\&
&\times[D(m^2-q^2+\lambda^2L)-2(D-2)Lq^2]exp[-T(m^2+\lambda^2L-q^2+ 
\frac{\alpha^2} {L+\alpha}q^2)]\nonumber\\& &=
-mg^2\frac{c_F}{(4\pi)^D}\mu^{4-D}\int_0^1d\alpha
\frac{D(m^2-\alpha^2)+2(D-2)(1-\alpha)q^2}{[m^2\alpha+(1-\alpha)
\lambda^2+q^2\alpha(1-\alpha)]^{2-D/2}}\Gamma(2-D/2).
\end{eqnarray}

As already mentioned, we shall refrain from working out the gamma-matrix
part of the propagator.
The point is that our methods have produced a one loop result in a
parametrized form pertaining to a configuration with an {\it open}
fermionic line, thereby extending the scope and use of the approach and
methodology of Ref. [7].

\vspace{1cm}

{\large {\bf 8. Concluding remarks}}

\vspace{.3cm} 

Our efforts, in this work, have centered around the role of the
spin factor in the worldline approach to gauge theories with
fermionic matter fields. Placing specific emphasis on open particle path
configurations we were able to accommodate  the spin attribute of the
propagating entity into a geometrical type of expression (spin factor) 
which registers as a
well defined weight factor in the path integral. Further to the point, its
presence facilitates the extraction of simple computational rules which,
at the perturbative level at least, produce algorithms of the Bern-Kosower
type [14] while extending, at the same time, their applicability to
fermionic Green's functions. On the other hand,
the dynamics of the particle-based casting of the system amounts to
emission and absorption of gauge field modes by the matter particle which 
is
solely controlled by a Wilson line operator. Our scheme thereby achieves a
neat separation between geometrical and dynamical aspects of physical
descriptions in gauge field theories. To summarize, our formulation has
not only made it possible to
accommodate  open fermionic paths, equivalently, Green's functions, into
worldline considerations but has produced computational rules that have 
clear as well as simple interpretations. 

The fact that our approach has
identified the Wilson line operator as the sole agent of the dynamics
operating in the system is worth some further comments. This occurrence
has proven to be of central importance to the dynamical
considerations we have been pursuing in connection to long distance
properties of QCD [10-12]. The wider realization that Wilson line
operators
carry the dynamics of eikonally-based considerations in QCD is, actually, 
well
founded in a variety of approaches which aim at factorizing soft
contributions to quantities of physical interest, such as form factors and
parton-parton scattering amplitudes [29-34]. A marked contrast is that in
the
worldline approach to QCD Wilson lines enter directly as an indigenous
component of the action whereas in standard field theoretical approaches
they are brought in, within the context of an operator formalism, for the
purpose of representing quarks as ordered
exponentials of the gluon field connected at the point where the hard
scattering occurs. In other words, they enter as sort of `tails' attached
to the quark field operator for the purpose of accounting for long
distance behavior. Our present work
further establishes the role of the Wilson line operators as the quantity
which carries the dynamics in a gauge field theory even beyond soft
effects.

Be that as it may, our immediate plans are to study situations wherein
external gluonic fields attach themselves to an open fermionic line
thereby extending the present considerations to vertex
functions and `Compton' amplitudes. We expect to report on progress along
these lines soon.  

\newpage

\appendix
\setcounter{section}{0}
\addtocounter{section}{1}
\section*{Appendix}
\setcounter{equation}{0}
\renewcommand{\theequation}{\thesection.\arabic{equation}}

\vspace{.3cm}

We display, below, some interim steps associated with the derivation of
formulas given in Section 7. Consider the particle-based action functional
$S$ as registered in eq. (79). Casting it in the form
\begin{equation}
S={1\over 4}\int_0^Td\tau\dot{x}^2(\tau)+{1\over
4L}\int_0^Td\sigma_1\int_0^Td\sigma_2B(\sigma_1,\sigma_2|\tau_1,\tau_2)
x(\sigma_1)\cdot x(\sigma_2), 
\end{equation}
with $ B(\sigma_1,\sigma_2|\tau_1,\tau_2)$ as given by eq. (82), we are
led to the equations of motion given by eq. (81) in the text.

Considering a solution of the form $x_\mu=\eta_\mu+z_\mu$, with $\eta_\mu$
as in (56) we are led to the relation (we drop, as in the text, the
$\tau$-arguments in $B$)
\begin{equation}
\int_0^Td\sigma'\left[\delta(\sigma-\sigma')\frac{\partial^2}{\partial 
\sigma'^2}-
{1\over L}B(\sigma',\sigma)\right]z_\mu(\sigma')= [\delta(\sigma-\tau_2)
-\delta(\sigma-\tau_1)](\tau_1-\tau_2)\frac{u_\mu}{L} 
\end{equation}
which justifies the introduction of $\Delta^{(1)}(\sigma,\sigma')$ according
to (83). One, now, readily determines
\begin{equation}
x_\mu^{cl}(\sigma)=\eta_\mu-\int_0^Td\sigma'\Delta^{(1)}(\sigma,\sigma')
[\delta(\sigma'-\tau_2)
-\delta(\sigma'-\tau_1)](\tau_1-\tau_2)\frac{u_\mu}{L}
\end{equation}
which produces eq. (84) in the text.

Employing the notation $\left[\delta(\sigma-\sigma')
\frac{\partial^2}{\partial\sigma'^2}-
{1\over L}B(\sigma',\sigma)\right]\equiv <\sigma|(\partial^2-{B\over
L})|\sigma'>$ we find, following the procedure applied in Refs [27,28] for
the closed contour case,
\begin{eqnarray}
& &-\Delta^{(1)}(\sigma,\sigma')= <\sigma|\left(\partial^2-{B\over
L}\right)^{-1}|\sigma'>=
 <\sigma|\left(1-{B\over L}\right)^{-1}\partial^{-2}|\sigma'>\nonumber\\
& &=<\sigma|\partial^{-2}|\sigma'> +{1\over
L}<\sigma|\partial^{-2}B\partial^{-2}|\sigma'> +{1\over
L^2}<\sigma|\partial^{-2}B\partial^{-2} B\partial^{-2}|\sigma'>+... 
\end{eqnarray}
with the first term in the series to be denoted, for the present purposes,
by $\Delta^{(0)}(\sigma,\sigma')$; it actually coincides with the free
bosonic propagator, cf. eq. (50).

Straight forward manipulations on the matrix elements entering
the above series result to the following, closed-form, relation
\begin{equation}
\Delta^{(1)}(\sigma,\sigma') =\Delta^{(0)}(\sigma,\sigma')- \frac
{[\Delta^{(0)}(\sigma,\tau_1)-\Delta^{(0)}(\sigma,\tau_2)]
[\Delta^{(0)}(\tau_1,\sigma')-\Delta^{(0)}(\tau_2,\sigma')]}
{L+G_B(\tau_1,\tau_2)}
\end{equation}

Liberal use of (A.3) and (A.5) leads to the relation
\begin{equation}
x_\mu^{cl}(\tau_2)-x_\mu^{cl}(\tau_1)=(\tau_2-\tau_1)\frac{L}{L+G_B(\tau_1,
\tau_2)}
\end{equation}
which, in turn, leads to eq. (89) in the text.

From (A.3) we obtain, after some straight forward manipulations, for
the derivative of the classical bosonic particle fields
\begin{equation}
\dot{x}_\mu^{cl}(\sigma)=u_\mu-{1\over 2}u_\mu(\tau_1-\tau_2)\frac
{\dot{G}_B(\sigma,\tau_1)-\dot{G}_B(\sigma,\tau_2)}{L+G_B(\tau_1,\tau_2)}
\end{equation}
which justifies eq. (88).

Turning our attention to expectation values, let us first furnish an
outline derivation of (90). Under the substitution $x\rightarrow x^{cl}+x$
we have that
\begin{equation}
<\dot{x}(\tau_1)\cdot\dot{x}(\tau_2)>=[
<\dot{x}(\tau_1)\cdot\dot{x}(\tau_2)>_o
+\dot{x}^{cl}(\tau_1)\cdot\dot{x}^{cl}(\tau_2)<1>_o]e^{-S^{cl}},
\end{equation}
where
\begin{equation}
<1>_o=\int_{x(0)=x(T)=0}{\cal D}x(\tau)exp\left[{1\over 4}\int_0^T
d\tau\dot{x}^2-{1\over 4L}
(x(\tau_2)-x(\tau_1))^2\right].
\end{equation}
One easily obtains
\begin{equation}
<1>_o=\frac{1}{(4\pi
T)^{D/2}}\left[\frac{L}{L+G_B(\tau_1,\tau_2)}\right]^{D/2}.
\end{equation}
On the other hand
\begin{eqnarray}
& &<\dot{x}(\tau_1)\cdot\dot{x}(\tau_2)>_o=-\frac{\delta^2}{\delta
J_\mu(\tau_1)
\delta J_\mu(\tau_2)} \int_{x(0)=x(T)=0}{\cal D}x(\tau)\nonumber\\&
&\quad\quad\quad\times
exp\left[{1\over 4}\int_0^T d\tau\dot{x}^2-{1\over 4L}\int_0^Td\sigma_1
\int_0^Td\sigma_2B(\sigma_1,\sigma_2)-i\int_0^Td\tau 
J(\tau)\cdot\dot{x}(\tau)
\right]_{J=0}.
\end{eqnarray}
Now, from (A.5) we determine
\begin{equation}
\frac{\partial^2}{\partial\sigma\partial\sigma'}\Delta^{(1)}(\sigma,\sigma')
={1\over 2}\ddot{G}_B(\sigma,\sigma')-{1\over 
4}\frac{[\dot{G}_B(\sigma,\tau_2)-\dot{G}_B(\sigma,\tau_1)]
[\dot{G}_B(\sigma',\tau_2)-\dot{G}_B(\sigma',\tau_1)]}{L+G_B(\tau_1,
\tau_2)}.
\end{equation}
which facilitates the carrying out of the (functional) Gaussian
integration in (A.11). One finds
\begin{eqnarray}
<\dot{x}(\tau_1)\cdot\dot{x}(\tau_2)>_o&=&{1\over (4\pi
T)^{D/2}}\left({L\over
L+G_B(\tau_1,\tau_2}\right)^{D/2}\nonumber\\& &\times D\left[\ddot{G}_B
(\tau_1,\tau_2)-{1\over
2}\frac{\dot{G}_B^2(\tau_1,\tau_2)}{L+G_B(\tau_1,\tau_2)}
\right]
\end{eqnarray}
as per eq. (90) in the text.

Finally, eq. (91) is deduced as follows
\begin{eqnarray}
<(x(\tau_2)-x(\tau_1)^2>_o&=&<(x^2(\tau_2)>_o+<(x^2(\tau_1)>_o-
2<(x(\tau_2)\cdot
(x(\tau_1)>_o\nonumber\\&=&2D[\Delta^{(1)}(\tau_2,\tau_2)
+\Delta^{(1)}(\tau_1,\tau_1)-2\Delta^{(1)}(\tau_1,\tau_2)]<1>_o\nonumber\\&
&=2D\frac{LG_B(\tau_1,\tau_2)}{L+G_B(\tau_1,\tau_2)}<1>_o.
\end{eqnarray}

\end{document}